\documentclass{aa}

\usepackage{graphicx}

\usepackage{amssymb}

\def\ion#1#2{{\rm #1}\,{\sc #2}}

\begin{document}
   \title{New detections of Mn, Ti and Mg in damped Ly$\alpha$ systems:
          Toward reconciling the dust/nucleosynthesis degeneracy
   \thanks{Based on the UVES observations collected during the ESO prog. ID No. 67.A-0022 
           at the VLT/Kueyen telescope, Paranal, Chile
   }}
   
   %\subtitle{}

   \author{M. Dessauges-Zavadsky
          \inst{1}$^,$\inst{2}
          J. X. Prochaska
          \inst{3,4}
          \and
          S. D'Odorico\inst{1}
          }

   \offprints{M. Dessauges-Zavadsky}

   \institute{European Southern Observatory, Karl-Schwarzschildstr. 2, D-85748 Garching 
              bei M\" unchen, Germany \\  
              \email{mdessaug@eso.org,sdodoric@eso.org}
              \and
              Observatoire de Gen\`eve, CH-1290 Sauverny, Switzerland
              \and
              The Observatories of the Carnegie Institute of Washington, 813 Santa Barbara Street, 
              Pasadena, CA 91101 \\       
              \email{xavier@ociw.edu}
              \and 
              Hubble Fellow
             }
   
   \date{Received ; accepted}

   \authorrunning{M. Dessauges-Zavadsky, J. X. Prochaska and S. D'Odorico}

   \titlerunning{Mn, Ti and Mg abundances in DLAs}

   \abstract{
We have combined new high resolution UVES-VLT observations of a sample of four damped Ly$\alpha$ 
systems (DLAs) at redshifts between $z_{\rm abs} = 1.7-2.5$ with the existing HIRES-Keck spectra 
to undertake a comprehensive study of their physical conditions and their abundances of up to 15 
elements. In this paper, we present abundance measurements for Mn, Ti and Mg which are among the 
first presented in the literature at these redshifts. We confirm the underabundance of Mn with 
respect to Fe as observed in lower redshift DLAs and a trend with decreasing metallicity very 
similar to the one observed in Galactic metal-poor stars. This agreement between the Mn/Fe ratios in 
DLAs and in Galactic stars suggests that the DLAs and the Milky Way share some similarities in their 
star formation histories. However, these similarities must be cautiously interpreted and investigated 
in light of all of the elements observed in the DLAs. We have obtained a first measurement and a 
significant upper limit of the Ti abundance at $z_{\rm abs} \sim 2$ from the \ion{Ti}{ii} lines at 
$\lambda_{\rm rest} > 3000$ \AA, and we discuss how the relative abundance of this highly depleted 
element can be used to prove unambiguously any enhancement of the abundance of the $\alpha$-elements 
relative to Fe-peak elements. We present the abundances of Mg for two DLAs in addition to the single 
Mg measurement existing in the literature. Contrary to the trend expected from differential 
depletion, the [Mg/Si,S] ratios tend to be over-solar. The effect is at the level of the measurement 
errors, but worth investigating in a larger sample because it could be suggestive of a peculiar 
nucleosynthesis effect.
   }

   \maketitle
%
%________________________________________________________________

\section{Introduction}

A measure of chemical enrichment in high redshift galaxies can be obtained through the study of 
absorption line systems in quasars, specifically via damped Ly$\alpha$ systems (DLAs). These 
systems with $N$(\ion{H}{i}) $> 2\times 10^{20}$ cm$^{-2}$ dominate the neutral hydrogen content of 
the Universe and are likely the protogalactic gas reservoirs for the formation of the majority of 
stars today (e.g. Wolfe et al. 1995). They provide the best opportunity to measure accurately the 
chemical abundances of many elements for a variety of galactic systems spanning a wide redshift 
interval. By comparing these abundance measurements with the chemical evolution models and abundance 
patterns of our Galaxy and nearby galaxies, one can infer information on the star formation history 
and galaxy evolution of these distant objects.

In particular, abundances of the $\alpha$-elements (e.g. Si, O, S, Ar) relative to the Fe-peak 
elements (e.g. Fe, Cr, Ni) are of great importance for defining the chemical evolution history. 
Being produced mainly by Type II and Type Ia supernovae (SNe) with relatively different timescales -- 
$< 2\times 10^7$ ($\alpha$-elements) and $10^8-10^9$ yrs (Fe-peak elements) -- respectively, the 
$\alpha$/Fe ratios are strongly dependent on the lifetimes of the element progenitor, whereas 
[Fe/H]\footnote{[X/H] $\equiv \log$[$N$(X)/$N$(H)]$_{\scriptsize{\textrm{DLA}}}$ 
$- \log$[$N$(X)/$N$(H)]$_{\odot}$.} depends on the star formation rate (Tinsley 1979; Matteucci 
2001). Therefore, the [$\alpha$/Fe] versus [Fe/H] relation is a strong function of the star 
formation history.
%
%_______________________________________________________________

\begin{table*}[t]
\begin{center}
\caption{Metal abundances} 
\label{}
\begin{tabular}{l c c c c c c c}
\hline
Quasar  & $z_{\rm abs}$  & [Fe/H] & [Si/Fe] & [Zn/Fe] & [Mn/Fe] & [Ti/Fe] & [Mg/Fe]       
\smallskip
\\     
\hline    
Q0100+13   & 2.309 & $-1.78\pm 0.08$ & $+0.35\pm 0.07$$^{\star}$ & $+0.25\pm 0.04$ & $-$   & $-$             & $+0.44\pm 0.19$ \\ 
Q1331+17   & 1.776 & $-2.01\pm 0.09$ & $+0.61\pm 0.03$ & $+0.75\pm 0.05$ & $-0.15\pm 0.04$ & $< -0.45$       & $+0.83\pm 0.22$ \\
Q2231$-$00 & 2.066 & $-1.20\pm 0.09$ & $+0.40\pm 0.05$ & $+0.41\pm 0.07$ & $-0.16\pm 0.10$ & $+0.70\pm 0.09$ & $-$ \\
Q2343+12   & 2.431 & $-1.19\pm 0.06$ & $+0.54\pm 0.05$ & $+0.62\pm 0.06$ & $-0.21\pm 0.05$ & $-$             & $-$ \\
\hline
\end{tabular}
\begin{minipage}{160mm}
\smallskip
$^{\star}$ The detected \ion{Si}{ii} lines are all saturated, so we report the \ion{S}{ii} column 
density. \\
Abundances relative to the solar values of Grevesse et al. (1996). \\
The [Fe/H] ratios correspond to the total metallicities, whereas the [X/Fe] ratios are computed 
by summing only the column densities of the components detected both in the \ion{X}{ii} and 
\ion{Fe}{ii} profiles. 
\end{minipage}
\end{center}
\vspace{-0.4cm}
\end{table*}
%
%_______________________________________________________________

Unfortunately this scheme to discriminate between different star formation histories has led 
to contradictory conclusions and interpretations of the observed relative abundances of the damped 
Ly$\alpha$ systems (e.g. Lu et al. 1996; Centuri\'on et al. 2000). The principle difficulty is to 
disentangle nucleosynthetic contributions from dust depletion effects. Because we are studying 
gas-phase abundances in DLAs, the measured abundances may not represent the intrinsic composition 
of the system if part of the elements is removed from the gas to the solid phase (Savage \& Sembach 
1996). Several pieces of evidence show that some dust is indeed present in DLAs with a dust-to-gas 
ratio between 2 to 25\% of the Galactic value (Pei et al. 1991; Vladilo 1998). The presence of dust 
hence implies a depletion of refractory elements (e.g. Si, Fe, Cr, Ni) preferentially incorporated 
into dust grains. 

In such a mixed situation (nucleosynthesis plus dust), the analysis of DLA chemical histories is 
severely limited by the low number of routinely observed elements: {\em $\alpha$-element} Si and 
{\em Fe-peak element} Fe. Occasionally the very important non-refractory {\em $\alpha$-elements} O, 
S and Ar (located in the Ly$\alpha$ forest) and more frequently the non-refractory {\em Fe-peak 
element} Zn are detected. We show in this paper that Mn, Ti and Mg -- Fe-peak and $\alpha$-elements 
respectively -- are additional elements which can provide important clues to the nature of the DLA 
star formation history. Well studied in Galactic stars (Nissen et al. 2000; Prochaska \& McWilliam 
2000; Chen et al. 2000), relatively little attention has been granted to these elements in DLAs 
due to their difficult detections (for Mn see Lu et al. 1996; Pettini et al. 2000; Ledoux et al. 2002 
and for Ti see Prochaska \& Wolfe 1999; 2002; Ledoux et al. 2002). Ratios of Ti and Mn relative to 
Fe are of special importance because differential depletion and nucleosynthesis tend to work in the 
opposite sense. Therefore, these ratios help alleviate the dust/nucleosynthesis degeneracy inherent 
to many other ratios (Prochaska \& Wolfe 2002). Meanwhile, Mg is a refractory $\alpha$-element and a 
comparison of its abundance with Si, S and O provides greater insight into the dust depletion 
pattern inherent to DLAs. 

We report here three new Mn measurements in DLAs at $z_{\rm abs} \sim 2$ in addition to the 
two previous measurements of Lu et al. (1996) and Ledoux et al. (2002) at similar redshifts, two 
new Mg measurements representing the only Mg measurements in addition to the single previous 
measurement of Srianand et al. (2000), and the first Ti measurements from \ion{Ti}{ii} lines at 
$\lambda_{\rm rest} > 3000$ \AA\ in DLAs at $z_{\rm abs} \sim 2$.

In Sect.~2 we briefly review our observations. In Sect.~3 we describe how we have succeeded to 
measure the column densities of the weak lines we are interested in and we show the results. In
Sect.~4, we highlight the importance of our measurements by discussing individually element by  
element.
%
%_______________________________________________________________

\section{Observations}

We used the unique capability of the Ultraviolet-Visual Echelle Spectrograph UVES (D'Odorico et al. 
2000) on the VLT 8.2m ESO telescope at Paranal, Chile, to obtain high resolution spectra in the UV 
$\lambda$ 3150--4500 \AA\ and in the near-IR $\lambda$ 6700--10000 \AA\ of four relatively bright 
quasars -- Q0100+13, Q1331+17, Q2231--00 and Q2343+12 -- with known $z_{\rm abs} = 1.7-2.5$ DLAs. 
Exposure times of 7200--15700 s have been obtained per object. Slit widths of 1 arcsec in the 
UV and of 0.9 arcsec in the near-IR were fixed throughout the observations with a CCD binning of 
$2\times 2$ resulting in a resolution of FWHM $\simeq 6.9$ km s$^{-1}$ and 6.4 km s$^{-1}$, 
respectively. The spectra were reduced using the UVES data reduction pipeline implemented in the ESO 
{\tt MIDAS} package (Ballester et al. 2000), and we made a systematic check of each step of the 
pipeline reduction to make sure of the best result. An average signal-to-noise ratio per pixel of 
$\sim 30$, 55 and 45 was achieved at $\lambda$ $\sim 3700$ \AA, 7500 \AA\ and 9000 \AA, respectively.

By combining these UVES-VLT spectra with the existing HIRES-Keck spectra obtained by Prochaska \& 
Wolfe (1999) we cover the total spectral range from 3150 to 10000 \AA\ for the four observed quasars 
and measure for the first time the column densities of over 15 ions and 12 elements for each of 
their intervening DLAs. Such a large amount of information allows one to constrain the dust and 
photoionization effects in the studied DLAs, and thus to infer the likely star formation history of 
the galaxies associated with these DLAs. The comprehensive analysis of the spectra and the full 
interpretation of the elemental abundances of the four DLAs will be presented in a future paper 
(Dessauges-Zavadsky et al. in prep.). Here we focus on the Mn, Ti and Mg measurements in the 
UVES-VLT spectra and on their importance to understand the intrinsic abundance pattern in DLAs.
%
%________________________________________________________________

\section{Data analysis and results}

The detection of \ion{Mn}{ii}, \ion{Ti}{ii} and \ion{Mg}{ii} lines is particularly challenging at 
high redshift because the \ion{Mn}{ii} lines, the strongest \ion{Ti}{ii} lines and the \ion{Mg}{ii} 
doublet are at high rest-wavelengths, $\lambda_{\rm rest} = 2576, 2594, 2606$ \AA, 
$\lambda_{\rm rest} > 3000$ \AA\ and $\lambda_{\rm rest} \approx 2800$ \AA, respectively. The Mg 
abundance measurement is hampered by an additional difficulty, the usual saturation of the doublet. 
The \ion{Mg}{ii} $\lambda$1239,1240 lines -- generally found in the red wing of the damping 
Ly$\alpha$ profile -- constitute the only possibility to measure the Mg abundance. Moreover, since 
all these lines are weak, \ion{Mn}{ii} and \ion{Ti}{ii} lines can be easily blended with the numerous 
telluric absorption lines present at the near-IR wavelengths, whereas \ion{Mg}{ii} lines can be 
contaminated by the Ly$\alpha$ forest. The presence and analysis of other and stronger metal lines 
in the same DLA hence play an important role in determining whether lines are blended and in 
distinguishing very weak features from the noise.

We used a $\chi^2$ minimization routine {\tt fitlyman} (Fontana \& Ballester 1995) in {\tt MIDAS} 
to fit Voigt profiles to the observed DLA absorption lines well described as a complex of components, 
each defined by a redshift $z$, a Doppler parameter $b$, a column density $N$ and the corresponding 
errors. Metal species with similar ionization potentials (e.g. neutral and singly ionized species) 
can be fitted using identical component fitting parameters, i.e. the same $b$ (assuming that 
the macroturbulent motions dominate over thermal broadening) and the same $z$ in the same component 
of all metal species and allowing for variations from metal species to metal species in $N$ only. 
We used relatively strong (but not saturated) lines to fix the component fitting parameters ($b$ 
and $z$), and excellent profile fits can then be achieved for weak lines and for lines from the
Ly$\alpha$ forest where the probability of blending is high by allowing only the column density to 
vary. This method is the most accurate one to measure the column densities of elements like Mn, Ti 
and Mg.

The results are summarized in Table~1 and the fitting solutions are shown in Figs.~1--4. We 
had a sufficient number of metal line profiles (on average 6 lines) to well constrain the fitting 
parameters in the studied DLAs exhibiting multicomponent velocity structures. The telluric 
lines have been identified thanks to the spectra of a hot, fast rotating star taken in the same 
nights as the scientific exposures. The \ion{Mg}{ii} $\lambda$1239,1240 lines detected in the DLA 
Ly$\alpha$ damping line red wing have been measured after a local renormalization of the spectrum 
around the \ion{Mg}{ii} lines with the fit of the Ly$\alpha$ damping wing profile. The errors on the 
measured Mg column densities have been estimated by varying the continuum level by 5\%.

The [Fe/H] ratios given in Table~1 correspond to the total metallicity measured in the DLA systems
obtained by summing the column densities of all the detected velocity components in the \ion{Fe}{ii} 
profiles. The [X/Fe] ratios with X = Si, Zn, Mn, Ti and Mg, on the other hand, were computed by 
summing only the column densities of the components detected both in the \ion{X}{ii} and  
\ion{Fe}{ii} profiles. In this way we avoid an overestimation of the abundance of Fe relative to 
other metal species obtained from weaker line transitions in which generally only the strongest 
components are detected. In the case of very weak lines, like the \ion{Ti}{ii} lines, one can indeed 
underestimate the [X/Fe] ratios by up to $0.3-0.4$ dex by considering the total Fe abundance. In the 
four studied DLAs this effect is particularly important in the DLAs toward Q2343+12 and Q2231-00 
which show complex metal line profiles with a large number of components. The same approach has 
already been used by Ledoux et al. (2002).

A component by component analysis of the metal ratios in individual DLA systems is interesting 
because it can reveal spatial fluctuations in the relative abundances (e.g. different chemical 
composition, different depletion level). We are not going to carry out this type of analysis in this 
paper, since, due to the weakness of the Mn, Ti and Mg lines, the results would not be reliable. We 
plan to discuss the possible component to component variations in a future paper where all 
detected element ratios will be presented (Dessauges-Zavadsky et al. in prep.).
%
%_______________________________________________________________

\begin{figure}[!]
\centering
   \includegraphics[width=9cm]{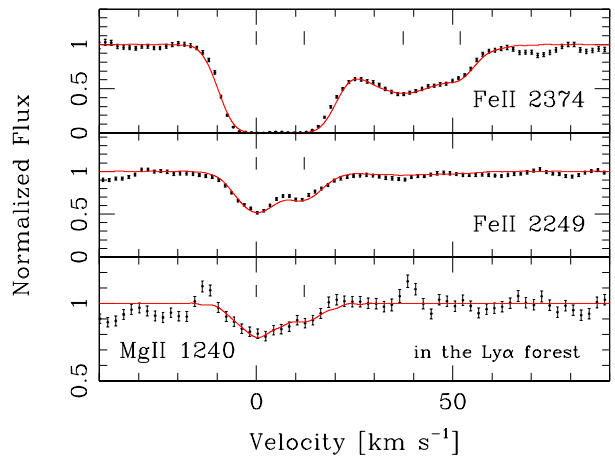}
\caption{Velocity plots of the metal line transitions (normalized intensities shown by dots with
1$\sigma$ error bars) for the DLA toward Q0100+13. The zero velocity is fixed at $z = 2.309027$. 
The vertical bars mark the positions of the fitted velocity components and the symbols $\oplus$ 
correspond to the telluric lines.}   
\label{}
\end{figure}
%
%________________________________________________________________

\subsection{Q0100+13, $z_{\rm abs} = 2.309$}

This system called also PHL 957 was carefully studied in Wolfe et al. (1994) and subsequently in
Prochaska \& Wolfe (1999) and Prochaska et al. (2001). We have obtained a very similar 
$\log N$(\ion{H}{i}) column density of $21.37\pm 0.08$. We confirm their column density measurement 
of $N$(Fe$^+$) obtained in the UVES spectra from the \ion{Fe}{ii} $\lambda$2249,2260,2344,2374 lines 
which were not present in the HIRES spectra. We also agree with their column density measurement of 
$N$(Zn$^+$). We now present a measurement of the Mg abundance from the \ion{Mg}{ii} $\lambda$1240 
line located in the red wing of the DLA Ly$\alpha$ profile, the \ion{Mg}{ii} $\lambda$1239 line being 
blended with Ly$\alpha$ clouds. Fig.~1 shows the \ion{Fe}{ii} and \ion{Mg}{ii} profiles where the 
local continuum around Mg$^+$ was defined with the fit of the damping profile of the Ly$\alpha$ line. 
The \ion{Mn}{ii} and \ion{Ti}{ii} lines in this DLA are outside the wavelength coverage of our 
spectra.
%
%_______________________________________________________________

\begin{figure}[t]
\centering
   \includegraphics[width=9cm]{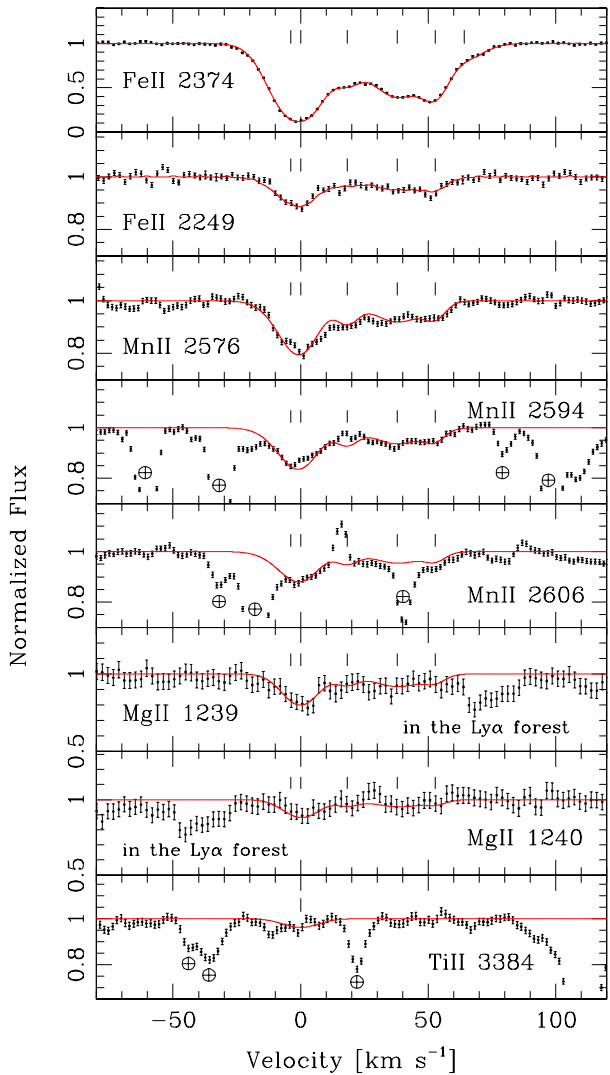}
\caption{Same as Fig.~1 for the DLA toward Q1331+17. The zero velocity is fixed at $z = 1.776370$.}
\label{}
\end{figure}
%
%_______________________________________________________________

\subsection{Q1331+17, $z_{\rm abs} = 1.776$}

Many analysis of the damped system toward the very bright quasar Q1331+17 (V=16.7) have been 
conducted. We mention here the two most accurate and recent ones of Prochaska \& Wolfe (1999) and 
Prochaska et al. (2001). In these two papers the authors have assumed the $N$(\ion{H}{i}) measured 
by Pettini et al. (1994) from relatively low-resolution spectra (1.3 \AA). We have obtained a new 
measurement of $\log N$(\ion{H}{i}) $= 21.14\pm 0.08$ from the UVES spectra and confirm that the DLA 
Ly$\alpha$ line is blended with another strong absorption system at $z_{\rm abs} = 1.786$ with 
$\log N$(\ion{H}{i}) $= 19.80\pm 0.10$. 

We agree with their Fe$^+$ column density measurement obtained in the UVES spectra from the 
\ion{Fe}{ii} $\lambda$2586 line not observed in the HIRES spectra. We also confirm the revised 
$N$(Zn$^+$) value of Prochaska et al. (2001) obtained by taking into account the contamination of 
\ion{Zn}{ii} $\lambda$2026 from the \ion{Mg}{i} $\lambda$2026 profile. The UVES spectra allowed us
to accurately measure this contamination thanks to the observations of the relatively strong line of 
\ion{Mg}{i} $\lambda$2852. With [Zn/Fe] $= 0.75\pm 0.05$, this system exhibits one of the largest 
Zn/Fe ratios (and hence dust depletion level) of any DLA.

We present here new measurements of some transitions not covered in the HIRES spectra, notably 
\ion{Mn}{ii} $\lambda$2576,2594,2606, \ion{Mg}{ii} $\lambda$1239,1240 and \ion{Ti}{ii} $\lambda$3384 
which is the strongest \ion{Ti}{ii} transition, shown in Fig.~2 with the \ion{Fe}{ii} profiles. 
The access to {\it three} \ion{Mn}{ii} lines has allowed us to obtain an accurate measurement of its 
abundance, although these lines are located in a relatively heavily contaminated area by telluric 
lines and, in addition, the \ion{Mn}{ii} $\lambda$2594 and 2606 lines are blended with some emission 
from not fully subtracted cosmic rays at $\sim 18$ and 15 km s$^{-1}$, respectively. The 
normalization in the region of the Mg doublet is not optimal over the entire range (specifically 
around the \ion{Mg}{ii} $\lambda$1239 line), but the adopted error on the Mg$^+$ column density takes 
this well into account. The detection of the \ion{Ti}{ii} $\lambda$3384 line is only marginal, and 
hence we adopt the measured Ti column density as an upper limit. The value we obtained is more 
constraining than the higher value deduced by Prochaska et al. (2001) from the \ion{Ti}{ii} 
$\lambda$1910 lines having a 3 times lower oscillator strength.
%
%________________________________________________________________

\begin{figure}[!]
\centering
   \includegraphics[width=9cm]{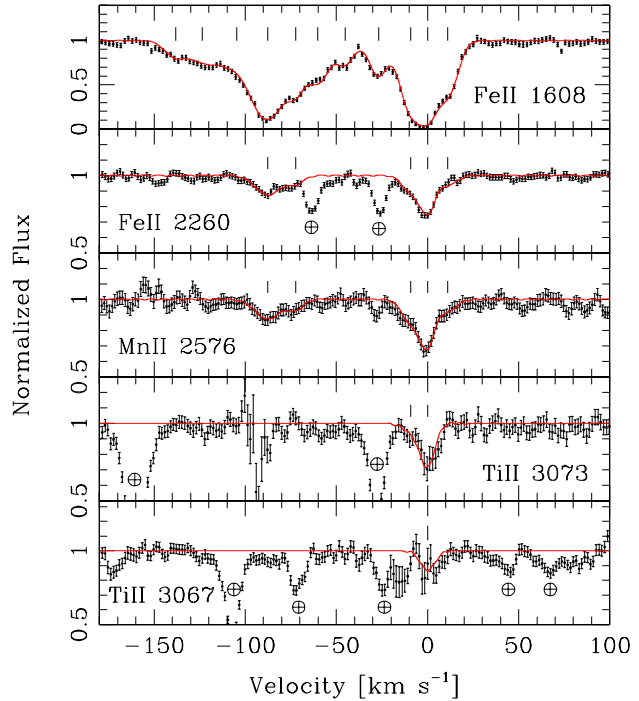}
\caption{Same as Fig.~1 for the DLA toward Q2231$-$00. The zero velocity is fixed at 
$z = 2.066161$.}
\label{}
\end{figure}
%
%_______________________________________________________________

\subsection{Q2231--00, $z_{\rm abs} = 2.066$}

This DLA system has previously been observed and analyzed by Lu et al. (1996) and Prochaska \& Wolfe
(1999), and recently completed by Prochaska et al. (2001). Throughout their analysis they have 
adopted the \ion{H}{i} column density measured by Lu \& Wolfe (1994). We have obtained here a new 
value for the $\log N$(\ion{H}{i}) of $20.53\pm 0.08$. We confirm their \ion{Fe}{ii} column density 
measurement and obtain an even more accurate measurement of $N$(Fe$^+$) by combining the \ion{Fe}{ii}
$\lambda$1608,1611 lines detected in the HIRES spectra and the \ion{Fe}{ii} $\lambda$2260,2344,2374 
lines observed in the UVES spectra. We also confirm their $N$(Zn$^+$) measurement. 

We now present a measurement of the Mn abundance and a new measurement of the Ti abundance obtained 
from the newly detected \ion{Ti}{ii} $\lambda$3067 and $\lambda$3073 lines. The \ion{Mg}{ii}
$\lambda$1239,1240 lines are unfortunately heavily blended with the Ly$\alpha$ lines. Fig.~3 shows 
the \ion{Mn}{ii} and \ion{Ti}{ii} profiles with the \ion{Fe}{ii} profiles. Only the \ion{Mn}{ii} 
$\lambda$2576 transition -- the strongest \ion{Mn}{ii} transition -- is covered in our UVES spectra. 
The anomalously strong feature at about $-30$ km s$^{-1}$ in the \ion{Mn}{ii} profile was not fitted 
because it is unlikely that it is an absorption from \ion{Mn}{ii}, since it appears to be at the 
noise level in this region of the spectrum (see for instance the features at the same intensities at 
$-140$, +50, +80 km s$^{-1}$). Moreover, other [X/Fe] ratios do not show any component at this 
velocity with a peculiar abundance behavior or dust depletion level relatively to the measured [X/Fe] 
ratios in the other velocity components (Dessauges-Zavadsky et al. in prep.). The measured value for 
the Ti$^+$ column density of $\log N$(\ion{Ti}{ii}) $= 12.66\pm 0.08$ is higher than the value 
obtained by Prochaska \& Wolfe (1999) from the \ion{Ti}{ii} $\lambda$1910 lines. Both the detection 
of Ti$^+$ and the measurement of N(Ti$^+$) presented here are more reliable and accurate, because 
the signal-to-noise in the UVES spectra is better and, in addition, the oscillator strength of the 
\ion{Ti}{ii} $\lambda$3073 line is slightly higher.
%
%________________________________________________________________

\begin{figure}[t]
\centering
   \includegraphics[width=9cm]{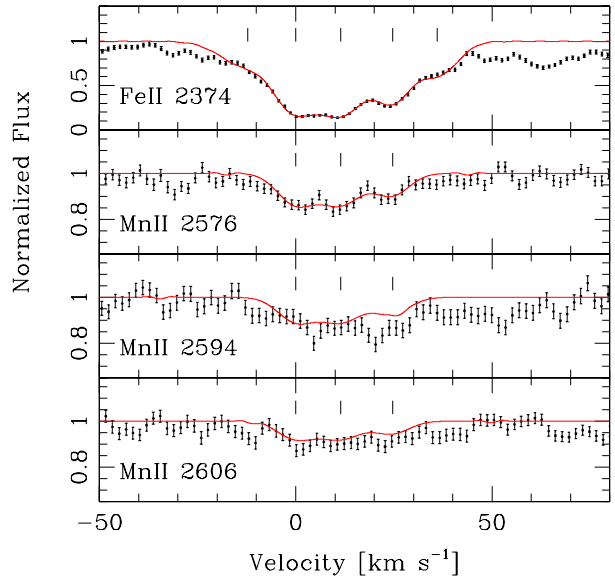}
\caption{Same as Fig.~1 for the DLA toward Q2343+12. The zero velocity is fixed at $z = 2.431157$.
Only a part of the metal line profiles extending over 300 km s$^{-1}$ is shown here, corresponding 
to the velocity interval where the weak lines are detected, i.e. the velocity interval of the 
strongest components.}
\label{}
\end{figure}
%
%________________________________________________________________

\begin{figure}[t]
\centering
   \includegraphics[width=7cm,angle=90]{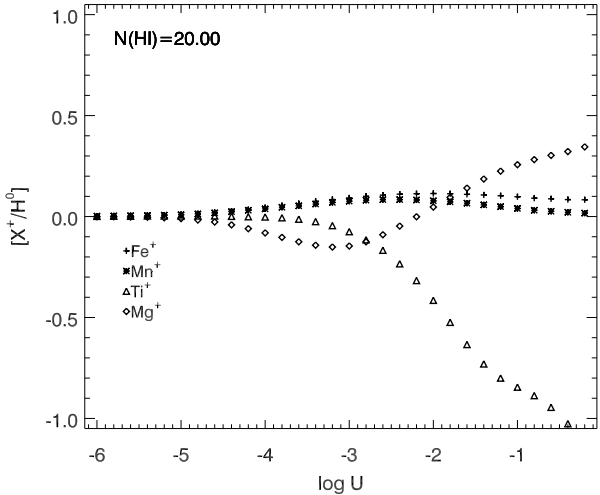}
\caption{Predicted logarithmic abundance of several ions X$^+$ relative to H$^0$ relative to the
intrinsic abundance of these two elements versus a range of ionization parameter U, assuming 
$\log N$(\ion{H}{i}) $= 20.0$, a metallicity [Fe/H] $=-1$, and the Haardt-Madau EUVB spectrum at 
$z = 2.5$. Departures of [X$^+$/H$^0$] from zero indicate that photoionization corrections are
required to calculate intrinsic elemental abundance ratios from the low-ion column densities.
Positive (negative) [X$^+$/H$^0$] values would imply overestimates (underestimates) of [X/H] from the
low-ion ratios. Photoionization corrections are small for all of the elements except Ti$^+$ which 
requires a large positive correction at larger U values.}
\label{}
\end{figure}
%
%________________________________________________________________

\subsection{Q2343+12, $z_{\rm abs} = 2.431$}

This system has never been previously studied in detail, only Lu et al. (1998) published the 
abundance measurements of some elements. We now present the Fe and Mn abundances, the \ion{Ti}{ii} 
and \ion{Mg}{ii} lines being outside the wavelength coverage of our UVES spectra, and in a future 
paper we will provide a complete analysis of all elements observed in this DLA (Dessauges-Zavadsky 
et al. in prep.). The absorption profiles of this DLA are characterized by a complex structure 
composed of numerous velocity components (22) and extended over 300 km s$^{-1}$ in velocity space. 
Fig.~4 shows the \ion{Fe}{ii} and \ion{Mn}{ii} profiles plotted over $\sim 80$ km s$^{-1}$ -- 
corresponding to the velocity interval of the strongest components and hence to the velocity interval 
where the weak lines, like the \ion{Mn}{ii} and the \ion{Zn}{ii} lines, are detected. The Fe$^+$ 
column density contained in this velocity interval corresponds to only $\sim 75$\% of the total 
$\log N$(\ion{Fe}{ii}) $= 14.66\pm 0.03$. This shows the importance when computing the [Zn/Fe] and 
[Mn/Fe] ratios (Table~1) to consider only the column densities of velocity components detected both 
in the \ion{Zn}{ii} and \ion{Fe}{ii} profiles and in the \ion{Mn}{ii} and \ion{Fe}{ii} profiles to 
avoid an underestimation of the abundances of elements obtained from weak lines relative to Fe.

The DLA Ly$\alpha$ line is outside our wavelength coverage, therefore we assume the 
$\log N$(\ion{H}{i}) $= 20.35\pm 0.05$ measured by D'Odorico et al. (2002). This \ion{H}{i} column 
density is relatively low and it is hence important and interesting to discuss the influence of 
ionization on the abundance pattern in this DLA. Here we present the effects of photionization on 
Mn, Ti and Mg relative to Fe for physical conditions relevant to the damped Ly$\alpha$ systems (see 
also Vladilo et al. 2001). Fig.~5 shows that for Mn and Mg the effects are small for nearly any 
assumptions on the ionizing radiation, but we find that one could underestimate Ti/Fe in some DLAs 
having a high ionizing parameter.
%
%________________________________________________________________

\section{Discussion}

In the following discussion we interpret our results on the Mn, Ti and Mg abundances and underline 
their complementarity to the measurements of other elements.
%
%_______________________________________________________________

{\em Manganese}. Major recent studies of Mn abundances in the Milky Way have been completed by 
Nissen et al. (2000) and Prochaska \& McWilliam (2000). About 120 [Mn/Fe] measurements from F and G 
stars in the metallicity interval $-1.4<$ [Fe/H] $<0.1$ have been obtained. Their results confirmed 
that Mn behaves in an opposite sense to the $\alpha$-elements: a steady decline to [Mn/Fe] $\sim 
-0.15$ from solar metallicity to [Fe/H] $\sim -0.7$ and a possible drop in [Mn/Fe] below [Fe/H] 
$\sim -0.7$. Mn is an Fe-peak element produced by massive stars and mostly by SNIa. Its behavior is 
not completely understood, although it is relatively well accounted for by metallicity-dependent 
yields for massive stars and for SNIa, illustrating a nice example of the ``odd-even'' effect for 
the Fe-nuclei (Goswami \& Prantzos 2000).

Although Mn is a refractory element, the Mn/Fe ratio is an important and useful diagnostic ratio.
Indeed, from Savage \& Sembach (1996) we can see that Mn and Fe have nearly identical depletion 
levels in the warm halo ISM clouds (the physical environment whose depletion level is most likely 
similar to the DLAs) and Mn is less depleted than Fe in the cold ISM clouds. Therefore, as stressed 
by Lu et al. (1996), sub-solar [Mn/Fe] ratios in gas-phase abundances must be interpreted in terms 
of nucleosynthesis while super-solar [Mn/Fe] ratios would require significant dust depletion. 
%
%________________________________________________________________

\begin{figure}[t]
\centering
   \includegraphics[width=9cm]{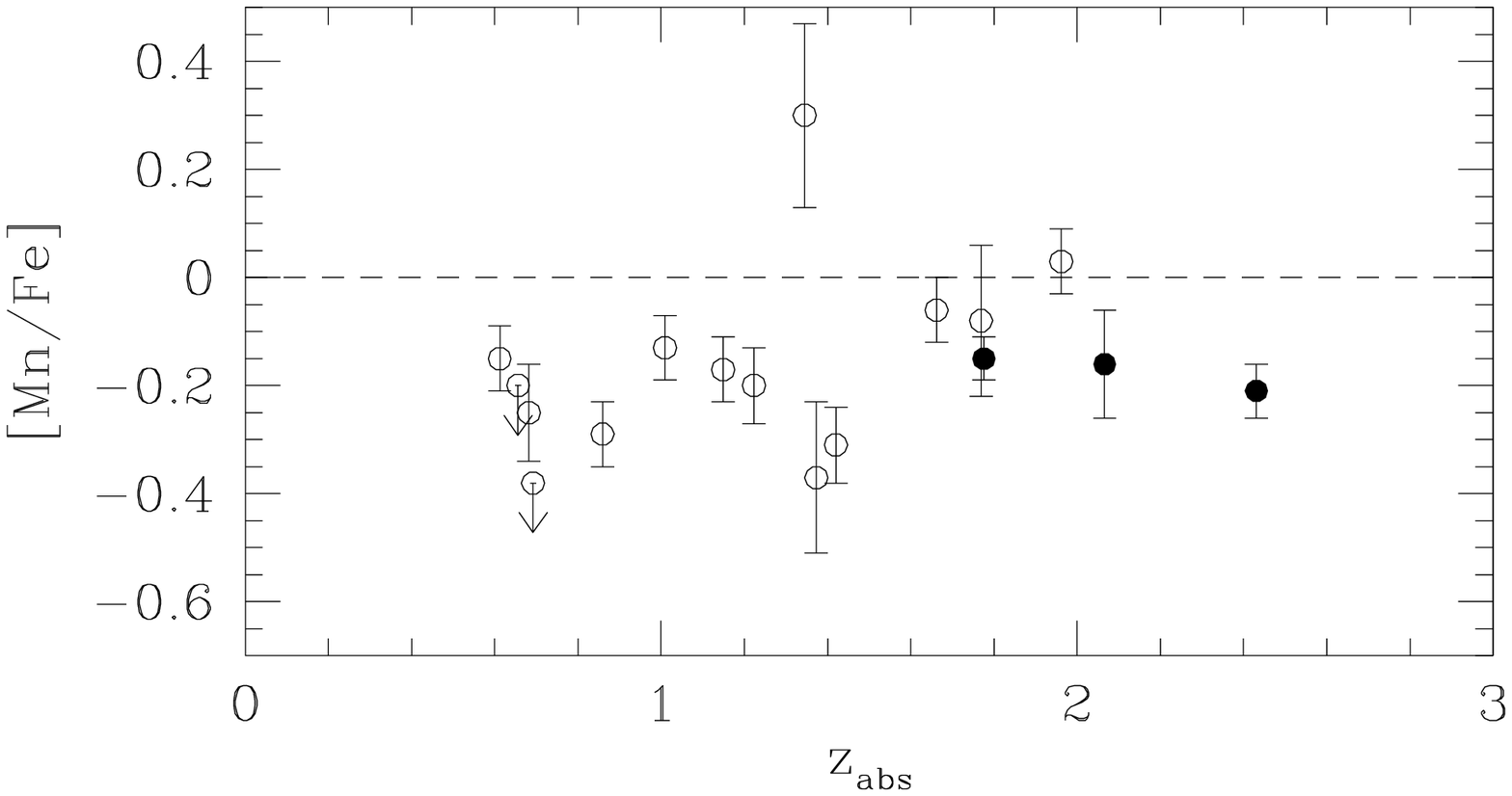}
   \includegraphics[width=9cm]{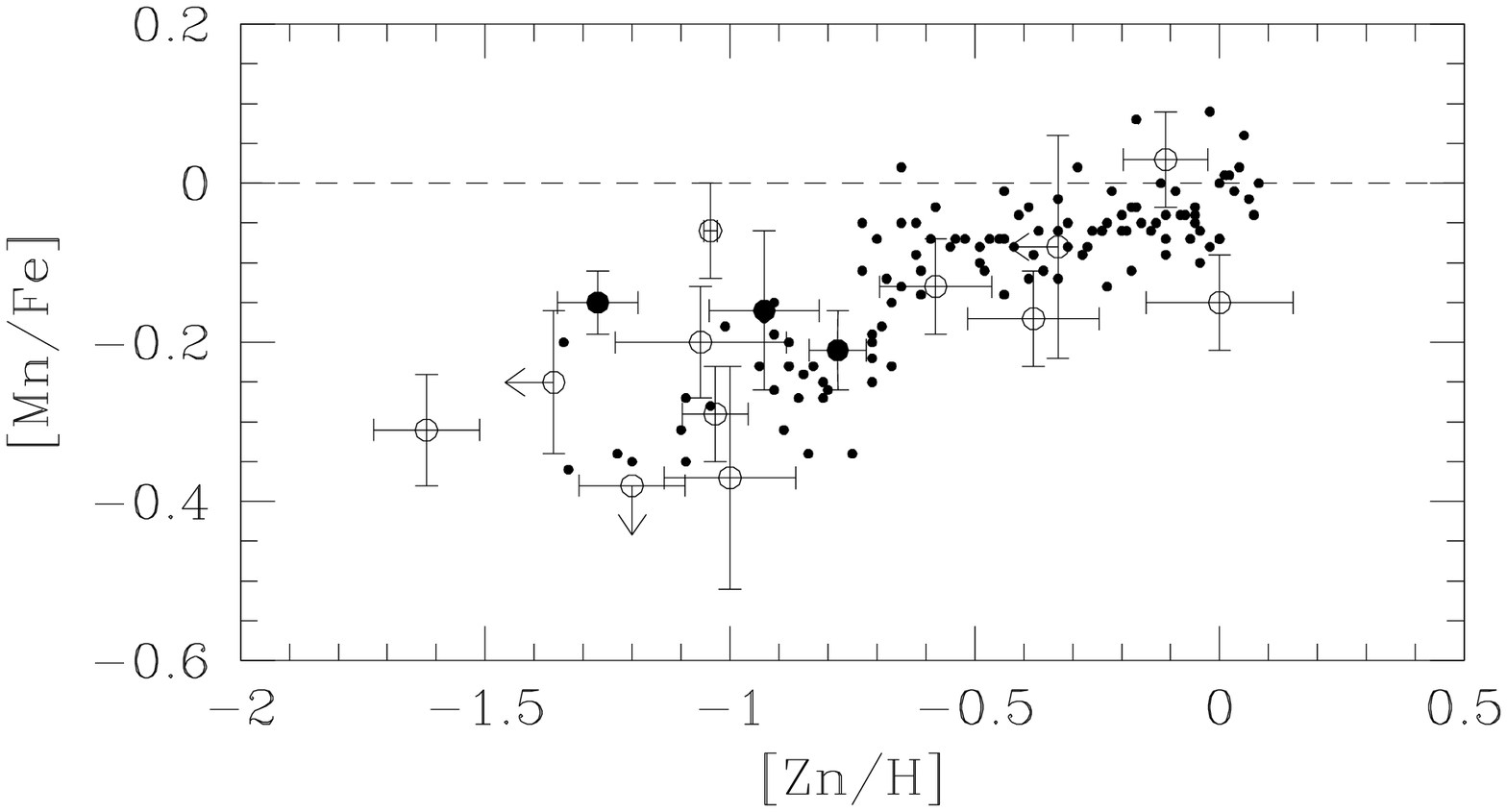}
\caption{Relative abundances of [Mn/Fe] versus $z_{\rm abs}$ of the DLA systems (upper panel) and 
versus the abundances of Zn, i.e. metallicities (lower panel). The open circles are literature 
values (Lu et al. 1996; Lopez et al. 1999; Pettini et al. 1999, 2000; Ledoux et al. 2002) which have 
been corrected from a possible underestimation due to a larger number of detected \ion{Fe}{ii} 
components relative to the \ion{Mn}{ii} components (see Section 3) when a detailed profile modeling 
was available. The filled circles are our measurements presented here. We have included to this 
sample only the DLAs for which a direct \ion{H}{i} column density measurement is known and we assume 
that in DLAs the metallicity [Zn/H] $\approx$ [Fe/H] in absence of dust. The dots are [Mn/Fe] versus 
[Fe/H] values measured in Galactic stars from Prochaska \& McWilliam (2000).}
\label{}
\end{figure}
%
%________________________________________________________________

The three new [Mn/Fe] measurements in the DLAs at $z_{\rm abs} \sim 2$ (see Table~1) are in 
agreement with the [Mn/Fe] measurements in DLAs at $z_{\rm abs} < 2$ (Lu et al. 1996; Lopez et al. 
1999; Pettini et al. 1999, 2000; Ledoux et al. 2002), and confirm that Mn is underabundant in 
galaxies associated with DLAs in a very similar way to Galactic stars (see Fig.~6). We also note 
that the DLA system toward Q1331+17 exhibits a much larger Mn/Fe ratio -- [Mn/Fe] $= -0.15\pm 0.04$ 
at a metallicity of [Zn/H] = $-1.27\pm 0.09$ ([Fe/H] $= -2.01\pm 0.09$) -- than the observed values 
in the Galactic stars at the same metallicity consistent with the fact that this DLA has a large 
depletion level.

The nice agreement between the Mn/Fe ratios in DLAs and in Galactic stars seen in Fig.~6 with a 
very comparable trend with decreasing metallicity, but a greater scatter in DLAs, suggests that the 
DLAs and the Milky Way share some similarities in their star formation histories. However, although 
the greater scatter may be explained by measurement uncertainties and the effects of differential 
depletion, these similarities must be cautiously interpreted and investigated in light of all of the 
elements observed in the DLAs. For example, it remains unclear what fraction of the damped systems 
exhibit $\alpha$-enhancements relative to Fe-peak elements consistent with the observations in 
Galactic metal-poor stars (e.g. Vladilo 1998; Centuri\'on et al. 2000; Prochaska \& Wolfe 2002). 
These similarities/discrepancies between the abundance patterns in DLAs and in Galactic stars have to 
be further carefully analyzed. They emphasize the importance of studying the Mn/Fe ratios in DLAs in 
addition to the $\alpha$/Fe ratios to better define the differences in the star formation histories 
of the DLAs and our Galaxy. Moreover, since direct observations of DLAs show a wide range of galactic 
morphological types (e.g. Le Brun et al. 1987), it is important to examine variations within their 
abundance patterns to search for variations in their star formation histories. Finally, by obtaining 
Mn measurements in low metallicity DLAs one can test nucleosynthetic models of Mn production -- 
major source of Mn: SNIa or SNII with strong metallicity dependence of the yield. 

{\em Titanium}. Many Ti stellar abundance measurements exist in our Galaxy over a wide metallicity 
interval $-4.0<$ [Fe/H] $<0.5$ (e.g. Chen et al. 2000). Ti is generally accepted as an 
$\alpha$-element, because it exhibits abundance trends similar to other $\alpha$-elements, although 
this behavior has not been reproduced by chemical evolution models (Timmes et al. 1995; Goswami \& 
Prantzos 2000). 

Thus far, only a few Ti measurements exist in DLAs at $z_{\rm abs} < 1.5$ (Ledoux et al. 2002), 
and efforts to measure Ti in DLAs at $z_{\rm abs} > 1.5$ have focused on the pair of very weak lines 
at $\lambda_{\rm rest} \approx 1910$ \AA\ and have yielded only a few tenuous detections (e.g. 
Prochaska \& Wolfe 1997, 1999; Prochaska et al. 2001). Our observations of the stronger \ion{Ti}{ii} 
$\lambda$3073,3242,3384 lines resulted in one robust detection and one significant upper limit.

Ti is a refractory element and exhibits an equivalent or higher depletion level in the ISM clouds 
than Fe (Howk et al. 1999). This runs contrary to the expectation for an $\alpha$-enhancement and, 
therefore, positive departures of [Ti/Fe] from the solar ratio is evidence for an 
$\alpha$-enhancement in the system independently of the presence of dust. Negative [Ti/Fe] ratios 
on the contrary should provide evidence for dust depletion. We found in the DLA toward Q2231$-$00, 
[Ti/Fe] $= +0.70\pm 0.09$ which suggests a dominance of Type II SNe relatively well consistent with 
the ratio of [S/Zn] $= 0.17\pm 0.09$ (Dessauges-Zavadsky et al. in prep.). In the DLA toward 
Q1331+17, on the other hand, [Ti/Fe] $< -0.45$ suggests significant dust depletion in agreement 
with [Zn/Fe] $= +0.75\pm 0.05$. 

{\em Magnesium}. Mg is an $\alpha$-element with a refractory nature whose depletion level is 
slightly higher than Si but lower than Fe (Savage \& Sembach 1996). Mg/Fe ratios are useful 
complements to other $\alpha$/Fe ratios measured in the same DLA to deduce information on the star 
formation history of the galaxy associated with the DLA. Moreover, the study of $\alpha$-element 
ratios with different depletion levels, like Si/Mg, Mg/Ti and Si/Ti, is another very interesting way 
to highlight the presence of dust in DLAs. But perhaps the most valuable aspect of the Mg 
measurements relates to the use of Mg equivalent widths to identify low redshift absorption galaxies 
when the Ly$\alpha$ lines are not detected (e.g. Bergeron \& Boiss\'e 1991; Rao \& Turnshek 2000). 
It is therefore important to know the range of Mg column densities in DLAs to confirm the empirical 
method to recognize them at low redshift.

While interesting for the reasons quoted above, the number of Mg measurements in DLAs remains tiny. 
With our two measurements a total of only three Mg abundance measurements are known in DLAs. They 
all show positive [Mg/S,Si] ratios of $0.10-0.20$~dex (see Table~1 and the DLA system in Srianand et 
al. 2000), in contradiction to the negative values expected from the differential depletion. But, 
since the detection of positive Mg/S,Si ratios is at the level of the measurement errors of 
$\sim 0.2$~dex and differs from the maximum over-depletion of Mg relatively to Si (of 0.2~dex) 
observed in the ISM warm gas by only 2$\sigma$ (Welty et al. 1999), it has to be further confirmed 
by additional Mg measurements before interpreting it as a peculiar nucleosynthetic effect similar to 
the observations of some Galactic stars (e.g. McWilliam et al. 1995).
%
%______________________________________________________________

\begin{acknowledgements}
       The authors wish to extend special thanks to all people working at ESO/Paranal for the high 
       quality of the UVES spectra obtained in service mode. M.D.-Z. is supported by an ESO 
       Studentship and the Swiss National Funds, and J.X.P. acknowledges support through the Hubble 
       Fellowship grant HF-01142.01-A awarded by STScI (NASA).
\end{acknowledgements}

\end{document}